\documentclass[doublecol]{epl2}
\usepackage{amsmath}
\usepackage{amsfonts}
\usepackage{mathrsfs}
\usepackage{epsfig}
\usepackage{graphicx}

\title{The strength of strong ties in scientific collaboration networks}

\author{Raj Kumar Pan\thanks{E-mail: \email{rajkumar.pan@aalto.fi}}
\and Jari Saram\"aki\thanks{E-mail: \email{jari.saramaki@aalto.fi}}}
\shortauthor{Raj Kumar Pan and Jari Saram\"aki}
\institute{%
BECS, Aalto University School of Science, P.O. Box 12200, FI-00076, Finland
}%

\date{\today}

\abstract{
  Network topology and its relationship to tie strengths 
  may hinder or enhance the spreading of information in social networks. 
  We study the correlations between tie strengths and topology in networks
  of scientific collaboration, and show that these are very different from ordinary
  social networks. For the latter, it has earlier been shown that strong ties are
associated with dense network neighborhoods, while weaker
  ties act as bridges between these. Because of this, weak links  act as bottlenecks for 
  the diffusion of information. We show that on the contrary, in co-authorship networks dense local 
  neighborhoods mainly consist of weak links, whereas strong links are
  more important for overall connectivity. The important role of strong links is 
  further highlighted in simulations of information spreading, where their
  topological position is
  seen to speed up spreading dynamics. Thus, in contrast to 
  ordinary social networks, weight-topology correlations enhance the flow
  of information across scientific collaboration networks.
 }


\begin{document}
\maketitle

\section{Introduction}


One of the key insights of network theory is that the structure
of networks reflects their function and it also sets constraints on dynamical 
processes taking place on networks~\cite{Newman06a}. Such 
structure may be a direct consequence of evolutionary forces acting on 
the entire system~\cite{Yamada2009,Pan07c}, such as for modules performing specific 
tasks in networks of metabolism or genetic regulation~\cite{Kashtan2005}. Alternatively,
the structure may arise in an emergent fashion from the actions of the 
individual nodes of the network. This is the case for social networks, 
where individuals attempt to satisfy their basic social needs related
to emotional support, social cohesion, and access to resources
and information, while
under spatial, time and cognitive constraints~\cite{Wasserman94,Granovetter73,
MoodyWhite2003,Dunbar1992,RobertsDunbarSN2009}.
In addition,
the evolution of networks of social interaction may be influenced by
 external driving 
forces; this is especially true for professional networks such as the 
networks of scientific collaboration considered in this Letter. 


Social networks are in general characterized by the existence of dense,
cohesive social groups that arise out of the above-mentioned individual-level 
mechanisms and constraints. A prominent mechanism giving rise
to dense social groups is triadic closure~\cite{Kossinets06,Kumpula07} --
learning to know people through the people we know. Simultaneously,
the interplay of several factors, such as homophily, where individuals 
of similar characteristics prefer to form ties~\cite{Kossinets06}, the
need for emotional support and social cohesion, and the high maintenance
costs of strong ties give rise to correlations between tie 
strengths and group structure. The existence of such correlations
was hypothesized by Granovetter~\cite{Granovetter73} already in the 1970's:
strong ties are associated with dense network neighborhoods, whereas
weak links act as bridges between these. This weak-link hypothesis
has since been confirmed with the help of electronic communication
records~\cite{Onnela07,Onnela07b,zhaoWL2010}. This particular
relationship between tie strengths and network structure has several
important consequences: first, for the connectivity of the entire network,
weak links play a crucial role~\cite{Csermely06}. Second, because of this, they also act
as bottlenecks for diffusion and spreading of information on the network.
When compared with a null model where tie strengths are replaced by
the network average, simulated spreading of information is slower~\cite{Onnela07}.


However, in networks of professional collaboration, such bottlenecks for information
diffusion would act against the purposes of individuals in the network. 
Whereas networks of scientific collaboration display many characteristic features of ordinary social 
networks, such as prominent community structure (see, e.g.,~\cite{newman2006,Palla2007,hou2008,LambiotteInformetrics2009}),
they are also shaped by different driving mechanisms. First, one can argue that the structure of the underlying space of ideas
and scientific knowledge are reflected in the network structure~\cite{Moody2004}. Second,
in addition to the need for cohesive sharing and processing of information in small groups, there is a particularly strong need for
avoiding scientific isolation by efficient transmission and brokerage of information in the network~\cite{Moody2004,Burt2004,LambiotteInformetrics2009}. 
These needs are likely to manifest in the network structure.
In this Letter, motivated by the above considerations and observations of anomalous weight-topology correlations in
collaboration networks~\cite{pan2011}, we show that unlike for ``everyday'' social networks, the correlations between tie strengths and network
topology enhance the spreading of information in networks
of scientific collaboration.

This paper is structured as follows: 
first, we describe the source data and characteristics of the co-authorship networks. 
Then, we address
correlations between link strengths and the surrounding network density,
and show that in scientific collaboration networks, dense network 
surroundings are associated with weak instead of strong links. We 
further corroborate this result by studying cliques, and show with
percolation analysis that strong links are more important to overall
connectivity. We then study the relationship of tie strengths to community
structure at several levels of coarse-graining. Finally, using simulated
spreading of information, we show that weight-topology correlations give
rise to fast spreading dynamics.

\section{Data Sets}
We consider two datasets: the first contains all articles published in the arXiv~\cite{Arxiv}
till March 2010 (595,276 papers), and the second
 all articles published in Physical Review (PR) journals~\cite{APS} between
1893-2009 (463,357 papers). From these data we extract the
list of authors,  identified by their surname and first two
initials. As our focus is on ties that have social aspects, we ignore
articles with $>10$ authors ($\sim$2$\%$ of all articles in each set)
 to filter out the huge collaborations in e.g. hep-ex and astro-ph,
where the number of authors can reach $\sim$1,000 and thus all authors are not
likely to know each other. We collapse the bipartite author-paper
networks to co-authorship networks by
connecting scientists who have co-authored one or more articles. 
We then extract the largest connected
components (arXiv: $N=$181,979 nodes and 
$L=$995,637 links, PR: $N=$203,245 nodes and $L=$1,198,002 links). These amount to 88.5\% and 94.4\% of
the total numbers of authors in the datasets, respectively.

For the tie strengths, \emph{i.e.}~link weights, of the unipartite projections
we use the formula introduced by Newman~\cite{Newman01a}: 
$w_{ij}= \sum_p \frac{1}{n_p - 1}$ where $p$ is the set of papers where
authors $i$ and $j$ collaborate and $n_p$ is the number of co-authors of paper $p$. 
Single-author papers are excluded. The motivation behind this commonly
used formula is that an author divides his/her time between the $n_p-1$ other
authors, and thus the strength of the connection should vary inversely with
$n_p-1$. It should be noted that this definition of tie strength is
not the only possible choice; however, it is in our view reasonable to assume
that joint work on a paper with a large number of authors contributes
less than, say, a two-author paper\footnote{For assessing the robustness
of our results, we have carried out similar analysis with an alternative
weighting scheme where $w_{ij}= \sum_p \frac{1}{(n_p -
1)^{\beta}}$. With $\beta=1$ we recover the original scheme, and 
with $\beta=0$ weights are insensitive to the number of authors of 
a paper. For $\beta=0.25$ and $\beta=0.5$, all our results hold,
while for $\beta=0$ resolution is lost for percolation and spreading
analysis, as 67$\%$ of the links have unit weight; nevertheless,
the rest of the results are qualitatively similar to the ones presented here.}.


\begin{figure}[t]
  \begin{center}
    \includegraphics[width=0.8\linewidth]{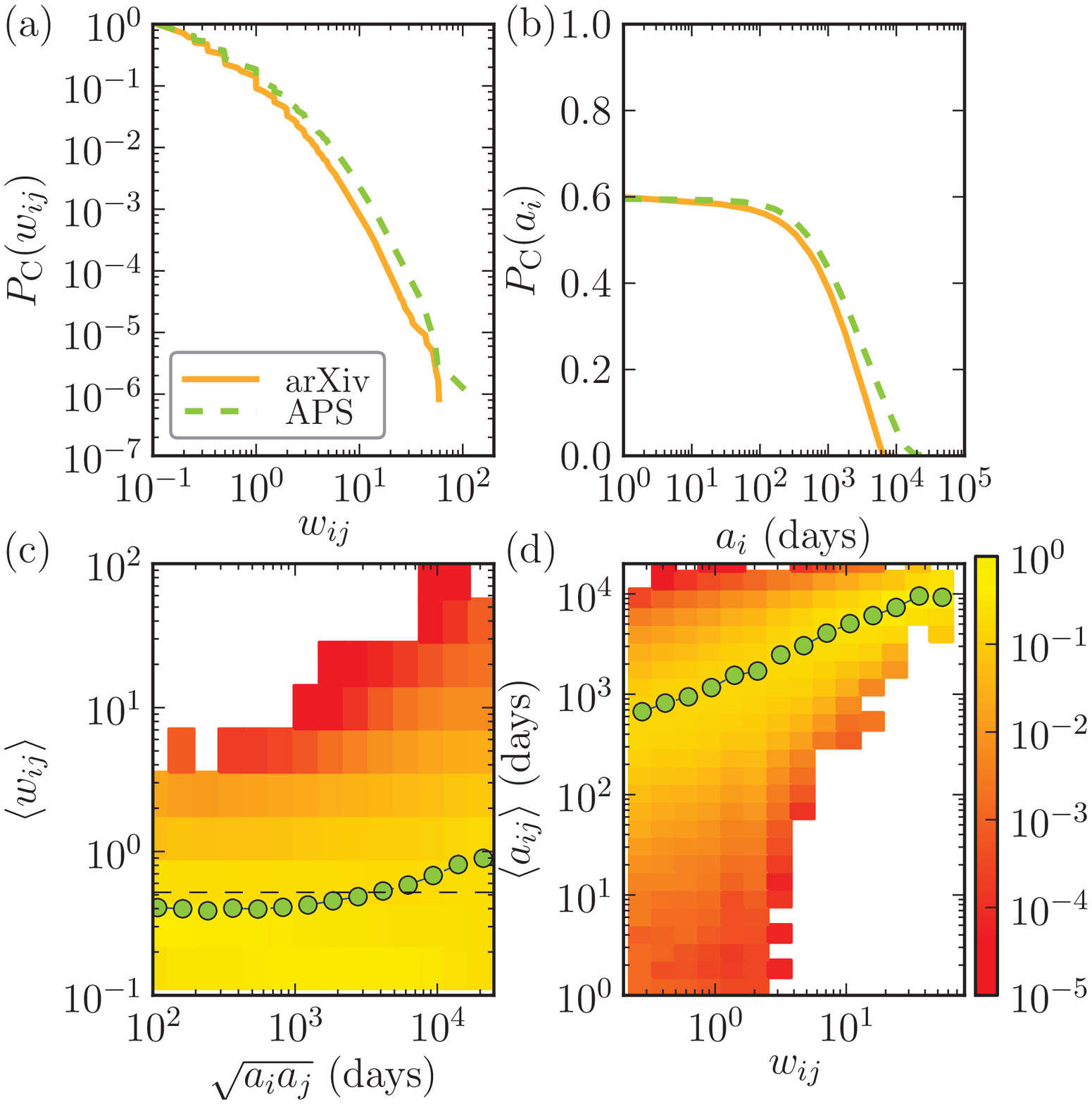}
  \end{center}
  \caption{Link and author statistics for the arXiv and PR networks.
   a) The cumulative link weight distributions. b) The cumulative distribution
   of the publication ages of authors (in days). c) The average link weight as a function of 
   the geometric mean of endpoint author ages for the APS network (circles). 
   The colors denote the corresponding vertically normalized probability
   distributions. d) The average link age in days as a function of the link weight
   for the APS network. The plots corresponding to panels c) and d) are 
   qualitatively similar for the arXiv network. }
  \label{fig:ageDist}
\end{figure}

\section{Results}
\subsection{Basic characteristics}
For both sets of data, the overall network properties are in accordance
with earlier observations~\cite{Newman01, Newman01a, Barrat04}:
the distributions of degree $k$  (number of links of a node) and strength $s$
(sum of link weights of a node) are heavy-tailed. Further, the strength approximately 
depends on degree as $\langle s \rangle \propto k\langle w \rangle$,
where $\langle w \rangle$ is the average link weight. The weight distribution
is also broad (Fig.~\ref{fig:ageDist} a). As high link weights are typically
accumulated over time between senior scientists, we define the 
publication age $a_i$ of  scientist  $i$ as the time elapsed between his/her first
and last publications in our records. Fig.~\ref{fig:ageDist} b) displays the 
cumulative distribution of such publication ages. Its shape confirms that most of
the scientists in the data can be considered junior, reflecting the hierarchy
of the scientific profession where the number of professors and other senior
scientists is significantly smaller than that of junior scientists.  Fig.~\ref{fig:ageDist} c)
displays average link weight as a function of the geometric mean of 
the publication ages of the endpoint authors; 
as expected, the link weights between senior scientists are on average higher . Similarly to the publication
age of scientists, we define the age of a co-authorship link $a_{ij}$ 
as the difference
between the dates of the last and first joint publications of the two authors $i$ and $j$.
As expected, this quantity increases on average with the weight of the link 
(Fig.~\ref{fig:ageDist} d). 

\begin{figure}[t]
\begin{center}
  \includegraphics[width=0.80\linewidth]{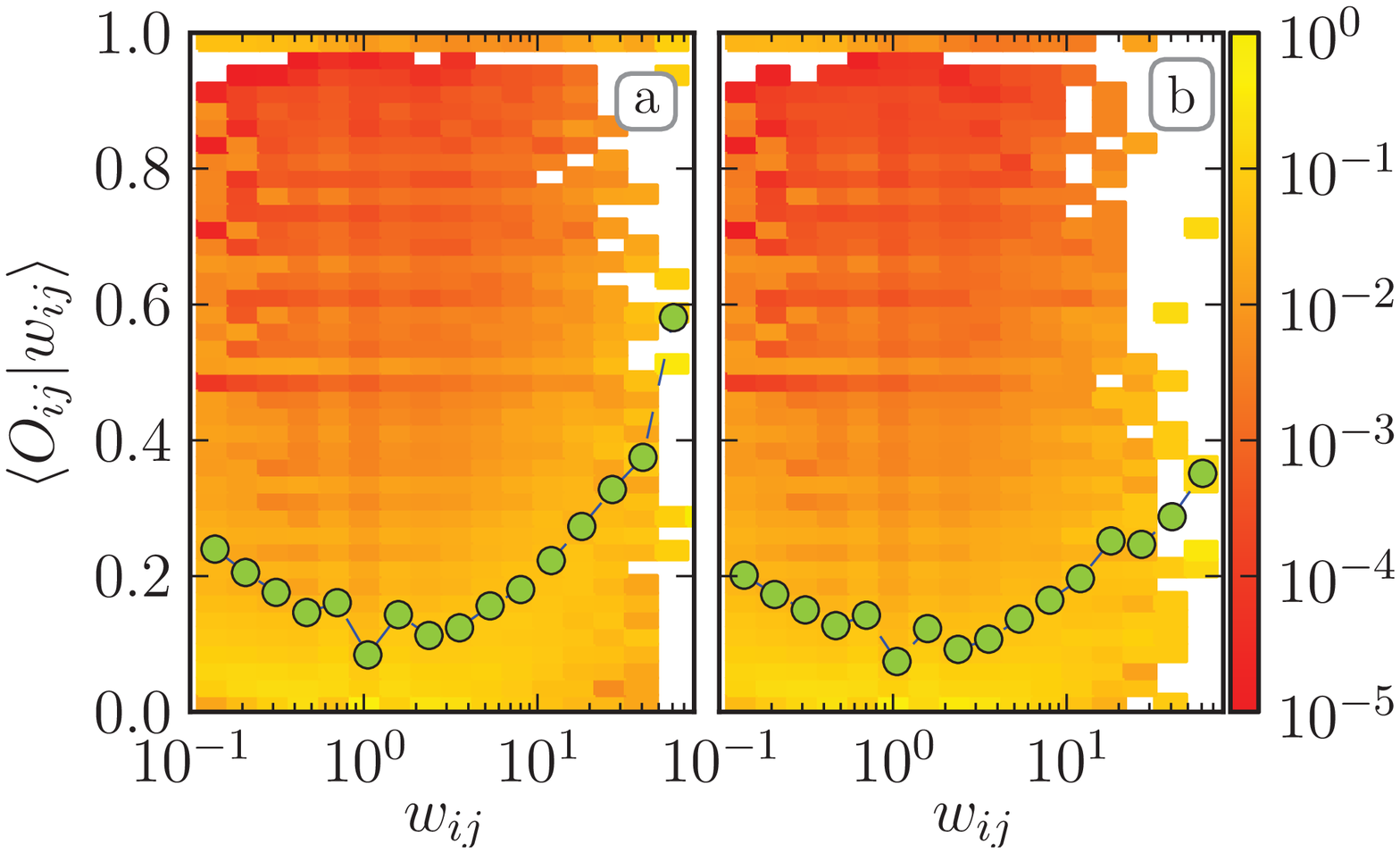}
\end{center}
\caption{The dependence of link overlap on link weight for (a) the arXiv and
(b) the PR co-authorship networks. Circles indicate averages, and colors
denote vertically normalized probability distributions.}
\label{fig:weightOverlap}
\end{figure}


\subsection{Dependence of neighborhood overlap on link weight}
We begin our exploration of the weight-topology correlations by 
considering the \emph{neighborhood overlap} of links. The overlap
$O_{ij}$ measures the fraction neighbors common to the endpoint
nodes of a link, and has earlier been observed to increase with link
weight in a communication network~\cite{Onnela07}, in accordance
with the Granovetter hypothesis~\cite{Granovetter73}. The overlap 
of a link is defined as
\begin{equation}
  O_{ij} = n_{ij}/(k_{i}-1+k_{j}-1-n_{ij}),
\end{equation}
where $n_{ij}$ is the number of neighbors common to the endpoint nodes $i$ and $j$
and $k_i$ and $k_j$ are their respective degrees.
In contrast to earlier results, we find that the overlap decreases with link weight
for both co-authorship networks (Fig.~\ref{fig:weightOverlap}) for the vast 
majority of links. This decrease is followed by an 
increase for the very highest-weight links in the tail of the weight distribution. Their number is very 
small: for the arXiv network, the section of the curve where
$w_{ij}>2$ only corresponds
to $\sim$ 5.3$\%$ of the links, and for the PR network, to $\sim$ 3.3 $\%$ of the links.
Hence, in stark contrast  to ordinary social networks, the weak links mainly reside inside
dense network neighborhoods, whereas strong links act as connectors between
these. Such weight-topology correlations reflect the hierarchy of the scientific
profession. As seen in Fig.~\ref{fig:ageDist} c) and d), weak links can
mainly be attributed to research groups that include junior scientists,
whereas strong links connect senior scientists
of different groups.
Further, the strongest links with high overlap belong to dense
neighborhoods, indicating long-term collaborations between senior scientists
of the same research group.

\subsection{Clique intensity distribution}

\begin{figure}
\begin{center}
  \includegraphics[width=0.80\linewidth]{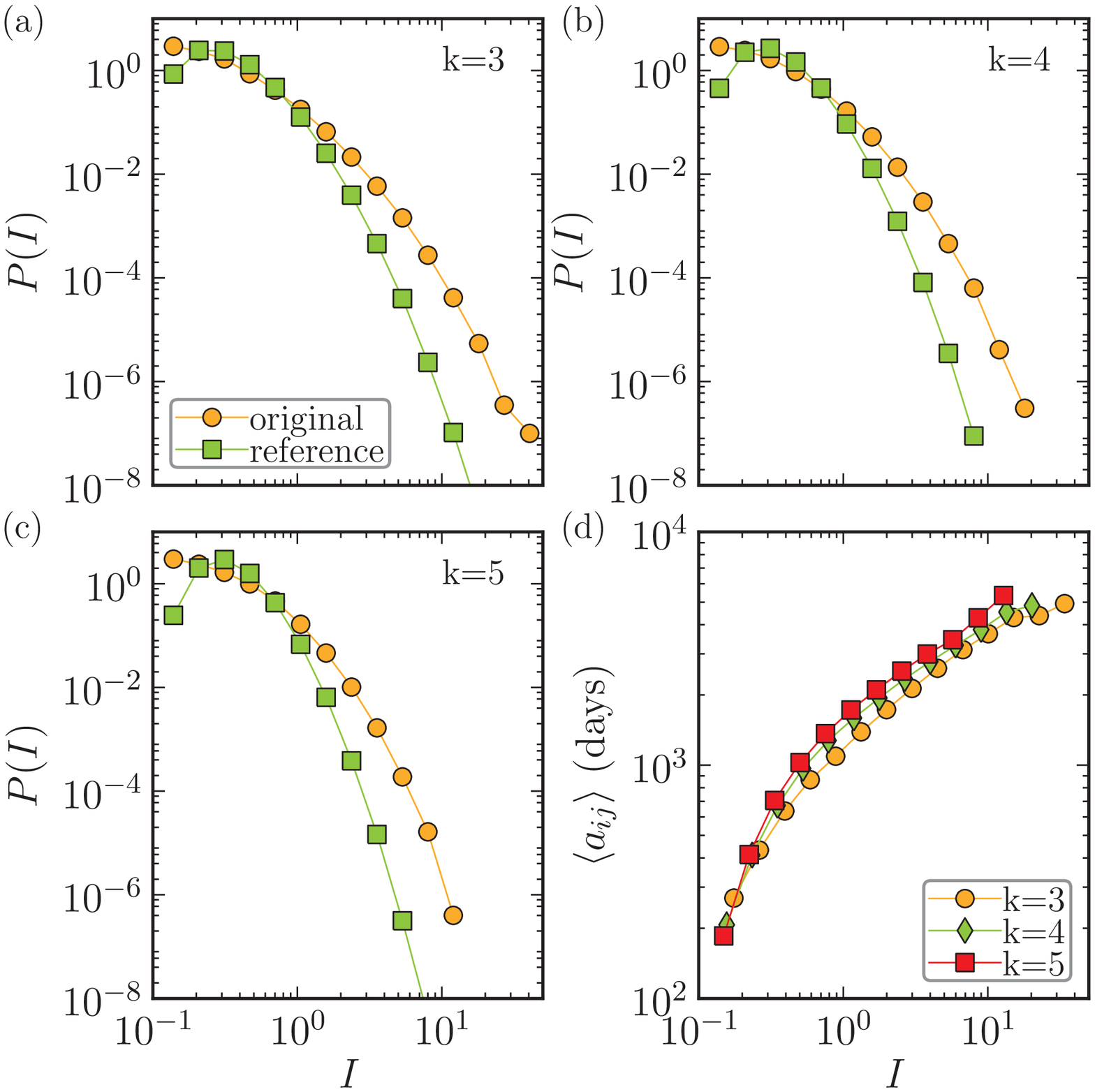}
\end{center}
\caption{a) to c) The probability distribution of the intensity of cliques of orders $k=$ 3,
4, 5 in the arXiv collaboration network and in the reference
ensemble. The reference ensemble is generated by shuffling the
weights of the empirical network while keeping its topology fixed; 
the reference distribution is an average over 100 realizations. 
The corresponding curves in the PR network are qualitatively similar. d)
The average publication age of the links of cliques as a function 
of their intensity.} 
\label{fig:intensity}
\end{figure}

The above results indicate that weak links are in general associated with
dense network neighborhoods. For  further evidence, we have
investigated subgraph weights by applying the concept of clique \emph{intensity}~\cite{Onnela05, Onnela07b},
designed for studying the coupling between the link weights and networks structure.
The intensity of a subgraph $g$ with nodes $v_g$
and links $l_g$ is given by the geometric mean of its weights as
\begin{equation}
  I(g) = \left[ \prod_{ij\in l_g} w_{ij} \right]^{1/|l_g|}
  \label{eqn:intensity}
\end{equation}
where $|l_g|$ is the number of links in $g$. 
Similarly to Ref.~\cite{Onnela07b}, we detect all $k$-cliques, that is, fully connected
subgraphs of $k$ nodes in the network, and calculate the distribution
of their intensities. As expected, the number of cliques of any order is much
larger than compared to a random configuration model with the same 
degree sequence. As a reference, we also calculate clique intensities
in an ensemble where the weights of the original network are 
randomly reshuffled, \emph{i.e.}~exchanged between its links, while 
the original topology and the number of cliques is retained.
Note that as the collaboration networks are projections
of bipartite networks, there is an abundance of cliques of
various sizes.
The intensity distributions of $k$-cliques for the original network and for
the reference ensemble are displayed in Fig.~\ref{fig:intensity} for the
arXiv network, with $k=3,4,5$.  First, we observe that in the original
network, the distribution of clique intensities is broad. There is a very
high number of low-intensity cliques, corroborating the overlap results.
This is in contrast with the results reported for the communication network
in Ref.~\cite{Onnela07b}, where the intensities are centered around a
well-defined mean. Overall, the median clique intensities in the reference
ensemble are larger than in the original networks (see
Table~\ref{tab:medianIntensity}).  The abundance of low-intensity cliques
is further highlighted when compared to the intensity distribution of the
reference ensemble. The broad distribution of the original network also
indicates that there is a small number of cliques with very high
intensities: as indicated in panel d) that shows the average publication
age of the links in cliques as a function of their intensity, such rare
high-intensity cliques correspond to strong collaborations between senior
scientists.

\begin{table}
  \begin{tabular}{|c|c|c|c|c|} \hline
   Order &\multicolumn{2}{c|}{arXiv} & \multicolumn{2}{c|}{PR} \\ \cline{2-5}
    $k$ & $I_{1/2}^{\mathrm O}$ & $I_{1/2}^{\mathrm R}$ & 
    $I_{1/2}^{\mathrm O}$ & $I_{1/2}^{\mathrm R}$ \\ \hline \hline
    3 & 0.330 & 0.347 ($10^{-17}$) & 0.297 & 0.312 ($10^{-4}$)\\ 
    4 & 0.323 & 0.354 ($10^{-5}$) & 0.303 & 0.316 ($10^{-4}$)\\
    5 & 0.326 & 0.357 ($10^{-4}$) & 0.327 & 0.318 ($10^{-5}$)\\
    6 & 0.331 & 0.358 ($10^{-4}$) & 0.401 & 0.320 ($10^{-5}$)\\ \hline
  \end{tabular}
  \caption{Median intensity $I_{1/2}$ of of cliques of order $k$, in the
  original arXiv and the PR collaboration networks (O) and the
  weight-shuffled reference ensembles (R). The order of magnitude of the standard
  deviation of the median across 100 realizations of the weight-shuffled reference ensemble is
  also indicated.}
  \label{tab:medianIntensity}
\end{table}

\begin{figure}[t]
\begin{center}
  \includegraphics[width=0.80\linewidth]{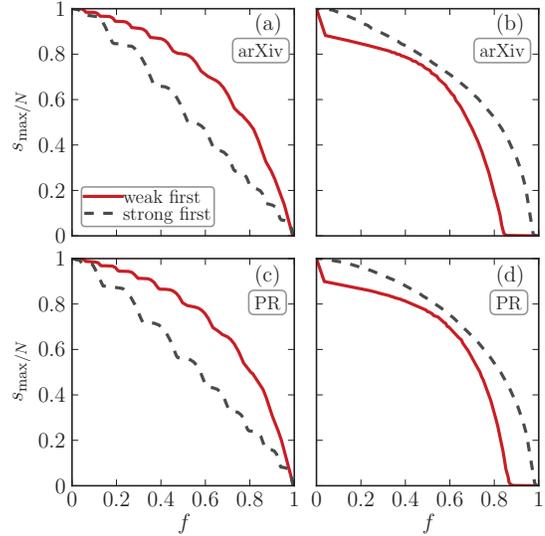}
\end{center}
\caption{Link percolation in the (a,b) arXiv and (c,d) PR
co-authorship networks. In both plots, the horizontal axis represents
the fraction of removed links $f$ and the vertical axis the relative
size of the giant connected component $s_{max}/N$.
Panels a) and c) show the dependence of the giant component size
on $f$ when links are removed in the order of increasing (solid lines)
and decreasing (dashed lines) weight. For both networks, removing
the strongest links first shrinks the size of the giant component fastest.
In panels b) and d), links are for reference removed in the order of 
increasing (solid lines) or decreasing (dashed lines) overlap; as expected, removing low-overlap
links first breaks the giant component faster.}
\label{fig:percolation}
\end{figure}

\subsection{Percolation analysis} \label{sec:percolation}

In order to understand the role of strong and weak links in the global
connectivity of the network, we next address link percolation in the
collaboration networks. Similarly to Ref.~\cite{Onnela07}, we 
first remove the links of the network in decreasing and increasing order
of weight, and keep track of the relative size of the largest connected component of nodes  $s_{\mathrm{max}}/N$ as a function of the fraction of removed links $f$.
The results are displayed in Fig.~\ref{fig:percolation}a) for the arXiv
and Fig.~\ref{fig:percolation}c) for the PR network. Both networks are remarkably
robust to link removal as the giant component only disappears when almost
all links have been removed, reflecting the broad degree distributions.
This is true for both orders of link removal. However, it is clear that the giant
component shrinks much faster when the strongest links are removed first,
indicating their important role for the overall connectivity of the network.
Again, this behavior is opposite to earlier observations~\cite{Onnela07,zhaoWL2010},
where the removal of weak links disrupts the connectivity faster.
We have also performed a similar analysis for 
overlap; it is seen that when removing low overlap
links first, the network fragments faster, and the giant component disappears
earlier than for weight removal (Figs.~\ref{fig:percolation} b) and d)). This behavior
can be attributed to modular structure, where the low-overlap links connect
dense regions of high-overlap links. 


\subsection{Modularity analysis}
To conclude our study of weight-topology correlations, we address
the mesoscopic structure of the co-authorship networks at different levels
of organization with community detection. The detection 
is based on the structure of the networks alone, 
\emph{i.e.}~unweighted links are used for detecting the communities, and the
relationship of link weights to the detected communities is then studied. 
In order to detect communities at different levels of coarse-graining,
we used the parametric generalization of modularity $Q$
introduced in Ref.~\cite{newman2006,Reichardt06} as
\begin{equation}
  Q_{\gamma}=\frac{1}{2} \sum_{i\neq j} \left(A_{ij}-\gamma P_{ij}\right)\delta_{c_i,c_j},  \label{eqn:modularity}
\end{equation}
where $A_{ij}=1$ if $i$ and $j$ are connected and 0 otherwise, $\gamma$ is the resolution parameter,
$P_{ij}=k_i k_j /2L$ represents the null model, and $\delta_{c_i,c_j}=1$ if the community
assignments $c_i$ and $c_j$ of the two nodes are the same. An optimal
partition corresponding to each value of $\gamma$ is obtained from maximizing the value of $Q_\gamma$.
The resolution parameter $\gamma$
allows for tuning the characteristic size of the modules. 
At small values of $\gamma$, large communities will be
detected. When $\gamma$ is increased, the optimization of $Q_{\gamma}$
leads to smaller and smaller communities in the optimal partition. We use
the Louvain method~\cite{Blondel08} to determine the optimal partition
corresponding to the maximum $Q_{\gamma}$.

\begin{figure}
\begin{center}
  \includegraphics[width=0.80\linewidth]{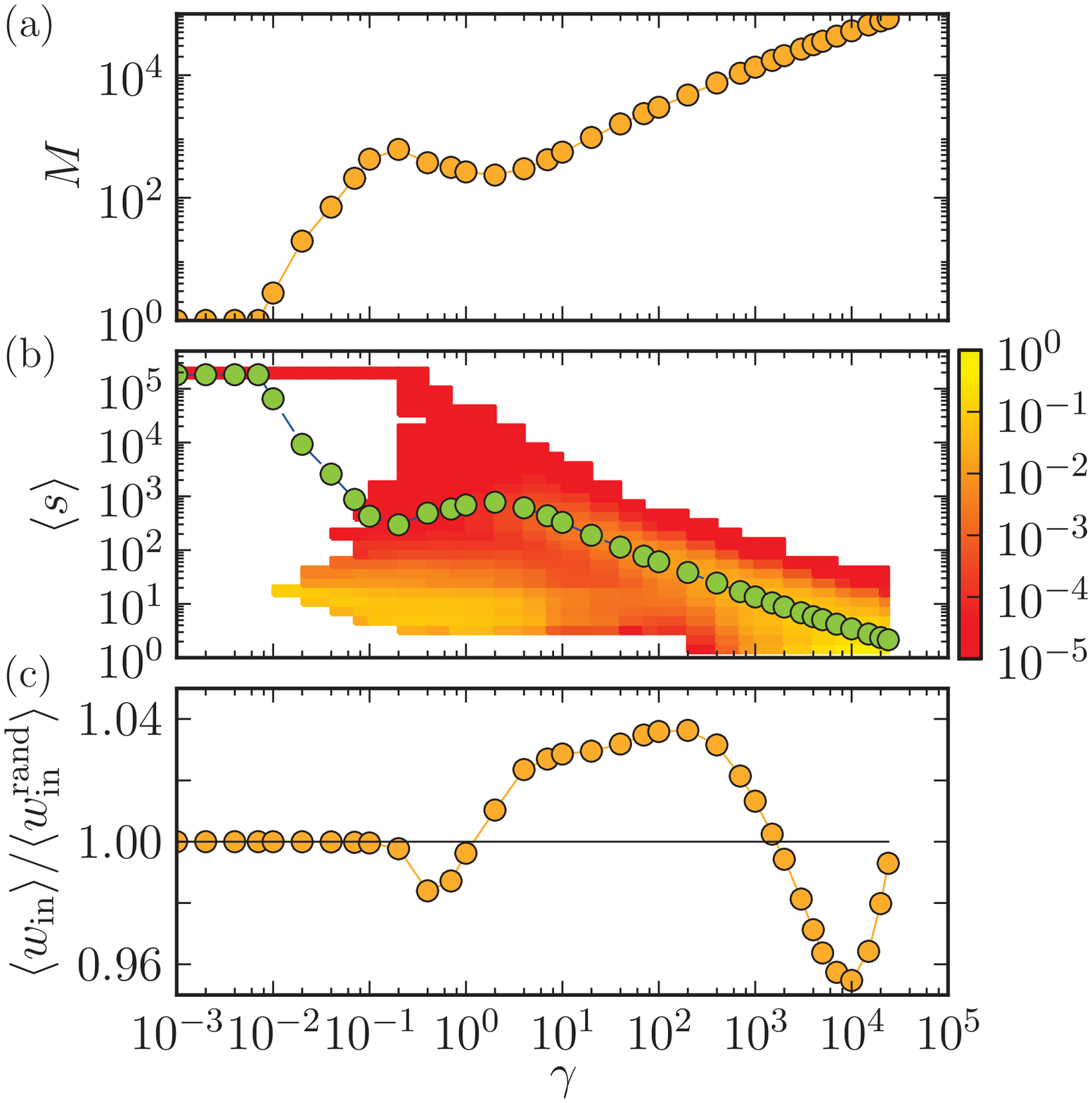}
\end{center}
\caption{Community structure in the arXiv network, at different levels of resolution. 
For all panels, the horizontal axis corresponds to the value of the resolution 
parameter $\gamma$, such that the resolution runs from coarse-grained to detailed. a) The number of detected communities.
b) The average size of detected communities (circles) and the corresponding vertically normalized
size distribution (colors). c) The average weight of community-internal links $\langle w_{\mathrm{in}}\rangle$, normalized by the 
average weight of the same links in the weight-shuffled reference ensemble $\langle w^\mathrm{rand}_\mathrm{in}\rangle$. The communities were
detected purely on the basis of topology, \emph{i.e.}~link weights were not
taken into account. Results for the  PR network are
qualitatively similar.} 
\label{fig:modularity}
\end{figure}

Fig.~\ref{fig:modularity} displays  the number of communities, their size,
and the average weight of their internal links relative to that in the
 weight-averaged reference ensemble, $\langle w_{\mathrm{in}}\rangle/\langle w^\mathrm{rand}_\mathrm{in}\rangle$,
 for different values of $\gamma$. 
For the smallest values of $\gamma$, the entire network is a single community. 
When $\gamma$ is increased, the method begins to pick up communities of 
fairly large size. For moderately large communities whose 
average sizes range from $\sim 10$ to $\sim 10^3$, 
it is seen that their internal link weights are higher than
randomly expected. This makes sense, as the sizes of the largest communities are 
of the order of fields or sub-fields of science.
As $\gamma$ is further increased
and the average community sizes drop below $10$, so that 
they are roughly in the range of research groups, intra-community
links have on average lower weights than randomly expected, in line with our
observations on the behavior of the link overlap and the analysis of cliques.
Overall, these results indicate that communities at different scales 
may display different weight-topology correlations.


\begin{figure}[ht!]
\begin{center}
  \includegraphics[width=0.80\linewidth]{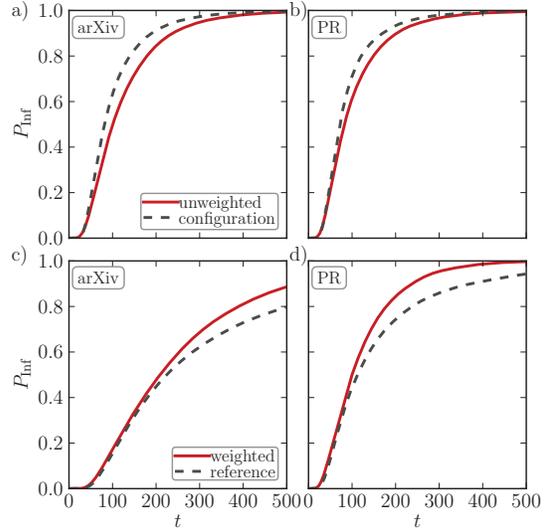}
\end{center}
\caption{SI spreading on the unweighted (a) arXiv and the (b) PR networks at
$p$=0.01  The spreading curves are compared with a randomized network, with
same degree distribution. SI spreading on the weighted (c) arXiv and (d) PR
networks. The spreading curves are compared with a randomized network, which
has same network structure but the weights on each of the edge is
shuffled.}
\label{fig:spreading}
\end{figure}

\subsection{Simulated spreading of information}
To conclude our investigations, we address the spreading of information in
the co-authorship networks, focusing on the role of structural correlations
and the relationship between tie strengths and topology\footnote{It is worth stressing that
we only address the spreading of information through weighted collaboration networks,
and use spreading dynamics as a probe of network structure. In reality,
the flow of information is of course not constrained to such idealized networks.}
Earlier, it has been shown with simulations that in social networks, the prominent
community structure, the effect of weak-link bottlenecks, and the
time-domain features of communication slow down spreading
 compared to randomized reference systems~\cite{Onnela07,Karsai11}. 
However, for the co-authorship networks, the weak-link bottlenecks appear
to be absent. To study the effects of network structure and weight-topology 
correlations on the spreading of information in the co-authorship networks,
we simulate spreading with the simple SI (Susceptible-Infectious) model. 
In this model, individuals are initially in the susceptible state (S), with the exception 
of a seed individual
whose state is set to infectious (I). The information then spreads through the
links of the network, such that at every time step, each susceptible individual
who is connected to an infectious individual becomes infected with some
probability $P_{ij}$ that may depend on the properties of the link
connecting the two nodes. 

Let us first study the effect of the network topology by disregarding weights
and setting $P_{ij}=p$ for all links. As a reference, we construct networks
where the degree sequence of the original networks is retained but links
are otherwise randomly rewired (the configuration model). This procedure
destroys structural correlations such as community structure. We then 
run the spreading simulation on the original and reference networks
by selecting random seed nodes, and observing the fraction of individuals
infected with the information $P_{Inf}$ as a function of time. Figs.~\ref{fig:spreading}
a) and b) show the resulting spreading dynamics on the arXiv and PR networks
and the corresponding reference ensembles, averaged over $10^3$ runs,
with $p=0.01$.  In both cases, it is seen that spreading is slightly 
slower in the original networks. This can be attributed to community
structure: the low numbers of links between communities slow
down spreading. 

However, when weights are introduced into the model, the situation is reversed.
For spreading on weighted networks, we set 
$P_{ij} = p \times w_{ij}$, \emph{i.e.}~the transmission rate between two
nodes is proportional to the link
weight $w_{ij}$. Here, the parameter $p$ now controls the overall spreading
rate. We set $p = 1 / \max(w_{ij})$, and so for the globally strongest 
link we have $P_{ij}=1$ and for others $P_{ij}<1$. To investigate 
the effect of weight-topology correlations, we apply the same reference
model as earlier, and randomly reshuffle link weights 
while keeping the network topology intact. We then simulate spreading as above.
Figs.~\ref{fig:spreading} 
c) and d) show that for both networks, the difference between original and reference
networks is the spreading is much faster in the original networks
compared to the reference, in contrary to the unweighted case. This effect could also reflect the existence of a core of
high-productivity scientists with strong ties, as observed in Ref.~\cite{Opsahl08}.
Hence, for the spreading of information, 
strong links and their position in the network are crucial in co-authorship networks.

\section{Conclusions and Discussion}
In conclusion, we have found that in networks of scientific collaboration,
the relationship between tie strengths and network topology is different
from ordinary social networks. This can be attributed to different driving
mechanisms of tie formation and reinforcement -- the strength of ties
reflects the hierarchy of the scientific profession as well as 
the needs of individuals, such as the need for efficient access to new information. 
Using neighborhood overlap and clique intensity analysis,
we have shown that locally dense network neighborhoods are associated
with weak links, whereas stronger links are of importance to the overall
connectivity of the networks. However, at a more coarse-grained resolution,
links within large communities of the size of fields or sub-fields of science
 are on average stronger than randomly expected. In future work,
 it would be interesting to explore these features deeper using information such
 as author departments and affiliations. This observation
 is also of importance for the design of community detection methods~\cite{Fortunato10}
 -- typically, weighted community detection methods assume that links within
 dense topological clusters are stronger than average, and our results
 indicate that this assumption is not necessarily valid. With the help of simulations, we 
 have also shown that the topological position of strong ties  in collaboration networks
increases the speed of spreading dynamics. Thus, weight-topology
correlations mitigate the isolating effects of small, cohesive groups and 
enhance the flow of information across
the network.

\section{Acknowledgments}
Financial support from EU's 7th Framework Program's FET-Open to
ICTeCollective project no. 238597 and by the Academy of Finland, the
Finnish Center of Excellence program 2006-2011, project no. 129670,  are
gratefully acknowledged.  We would like to thank Hang-Hyun Jo for useful
discussions.
\bibliography{coauthorship}


\end{document}